# Influence of Pressure on the Structure and Electronic Properties of the Layered Superconductor Y$_2$C$_2$I$_2$

**Kyungsoo Ahn† ‡  Reinhard K. Kremer‡ §, Arndt Simon‡ William G. Marshall£ , Peter Puschnig¶ , and Claudia Ambrosch-Draxl¶**

† Department of Chemistry, Yonsei University, Wonju 220-710, Korea
‡ Max-Planck-Institut für Festkörperforschung, Heisenbergstr. 1, D-70569 Stuttgart, Germany
£ ISIS Facility, Rutherford Appleton Laboratory, Chilton, Didcot OX11 0QX, United Kingdom
¶ Institut für Theoretische Physik, Universität Graz, Universitätsplatz 5, A-8010 Graz, Austria

**Abstract.**
 The structural properties under hydrostatic pressure up to 3 GPa of the layered rare earth carbide halide superconductor Y$_2$C$_2$I$_2$ are studied by neutron powder diffraction at room temperature. The compressibilities are anisotropic, such that the compressibility perpendicular to the layers being approximately twice as large as within the layers. The atomic positional parameters determined from the powder diffraction patterns as a function of pressure were used as the basis for highly resolved electronic structure calculations. These reveal the electronic density of states at $E_{\rm F}$ to increase with pressure. As is shown quantitatively, this effect outweighs the pressure induced lattice stiffening effects and is responsible for the pressure induced increase and saturation of $T_{\rm c}$ towards 2 GPa.

§ To whom correspondence should be addressed (rekre@fkf.mpg.de)

## 1. Introduction

It is widely accepted that layered materials provide favourable conditions for superconductivity with high $T_c$'s [1, 2, 3]. Well known examples which support this conjecture are the high-$T_c$ oxocuprate superconductors and more recently the 40 K superconductor $MgB_2$ [4]. Other superconductors with quasi two–dimensional structures are the transition metal dichalcogenides as well as organic systems which, however, have less spectacular transition temperatures. A central issue for superconductivity in layered materials is the question how the superconducting properties depend on the interlayer coupling. Interlayer coupling can favourably be tuned and superconducting properties eventually optimized by applying external pressure. Compression of the lattice, at first hand, leads to increased electronic hybridization with increased electronic bandwidth which may add to optimize the electronic conditions for superconductivity. On the other hand, pressure induced phonon hardening, in general, may dominate electronic effects. For example, the highest $T_c$ of 164 K among the superconducting oxocuprates was obtained in $HgCa_2Ba_2Cu_3O_{8+\delta}$ by applying a hydrostatic pressure of 30 GPa [5]. In contrast, pressure leads to a reduction of $T_c$ in $MgB_2$ and the increase of the characteristic phonon frequency under pressure has been identified as the dominant contribution [6, 7].

Here we report about a pressure study of the structural and electronic properties of the quasi-two-dimensional superconductor $Y_2C_2I_2$ ($T_c$=10 K)[8]. $Y_2C_2I_2$ belongs to the family of carbide halides of the rare-earth metals, $RE_2C_2X_2$ (X = Cl, Br, I and RE being a rare-earth metal) which crystallize with layered structures. These contain doublelayers of close-packed RE metal atoms with $C_2$ units occupying octahedrally coordinated metal atom voids [9, 10]. The doublelayers are sandwiched by layers of halogen atoms to form X–RE–$C_2$–RE–X slabs which connect in stacks along the crystallographic c-axis (cf. Fig. 1). The highest $T_c$ of 11.6 K has been reached by optimizing the electronic conditions using an appropriate mixture of halides in the compound $Y_2C_2Br_{0.5}I_{1.5}$[11]. A first study of the pressure dependence of $T_c$ of several $RE_2C_2X_2$ (RE=La, Y; X=Br,I) revealed a situation similar to the ambivalent behaviour of the transition metal elements [12, 13]: The variation of $T_c$ with pressure is markedly different for the $RE_2C_2X_2$ compounds, not only in magnitude but also in sign: While a positive increase of 18% and 10% for $dlnT_c/dP$ has been observed for $La_2C_2Br_2$ and $Y_2C_2I_2$, respectively, $T_c$ decreases with pressure for $Y_2C_2Br_2$ and $Y_2C_2Br_{0.5}I_{1.5}$. The magnitude of the decrease amounts to about 3–4%. This finding indicates that pressure optimizes the electronic conditions in $La_2C_2Br_2$ and $Y_2C_2I_2$ and this effect is able to outweigh the decrease of $T_c$ due to phonon hardening induced by the lattice compression. Since the compressibility perpendicular to the layers may be expected to be significantly larger than within the layers, these results also give a hint that tuning the interlayer distance is more essential for the electronic properties and superconductivity of these systems than hitherto considered.

Initially, band structure calculations have been carried out for $Y_2C_2Br_2$ based on the tight-binding linear muffin tin orbital method [11]. These revealed a peak in the electronic density of states (DOS) slightly below the Fermi level. This finding was experimentally supported by a nonlinear temperature dependence of the Korringa relaxation observed in $^{13}$C NMR experiments [14]. It was suggested that within a rigid band scenario the $T_c$ variation of the Cl/Br/I mixed compounds and the pressure experiments can be understood as a result of the Fermi level shifting through this peak in the DOS [12]. More recent electronic structure calculations performed for $Y_2C_2Br_2$ and $Y_2C_2I_2$ within the full-potential linearized augmented plane wave (LAPW) method essentially confirm the peak structure at approximately 0.1 eV below the Fermi level. There is, however, a qualitative change in the DOS close to the Fermi level rather than a simple volume effect when Br is replaced by I. The single peak in the DOS of $Y_2C_2Br_2$ splits into two for $Y_2C_2I_2$, and the Fermi level is located in the valley between these two peaks [15]. A similar scenario has been observed for $La_2C_2I_2$ for which extended Hückel band structure calculations find the Fermi level also in a local minimum of a pseudogap-like DOS [16]. The subtle but qualitative difference in the structure of the DOS near the Fermi level for the two systems $Y_2C_2I_2$ and $Y_2C_2Br_2$ strongly suggests that the assumption of rigid bands used to explain the Cl/Br/I substitution experiments has to be questioned. To explore the electronic structure near the Fermi level and to correlate the changes in the electronic structure with the superconducting properties we carried out a detailed study of the pressure dependence of the crystal structure of $Y_2C_2I_2$. The atomic parameters refined from pressure dependent neutron diffraction patterns are taken as input for highly resolved electronic structure calculations. These reveal a closing of the pseudogap in the DOS and a concomitant increase of the electronic density of states at the Fermi energy with pressure as the origin for the observed increase of $T_c$ with pressure. A detailed analysis of this effect and the counteracting lattice stiffening effects shows quantitative agreement with the experimental observations.

## 2. Experimental

Polycrystalline samples of $Y_2C_2I_2$ were prepared from Y, $YI_3$ and carbon powder as described in detail elsewhere [16]. The powders were pressed to pellets and annealed overnight at 1300 K in sealed Ta tubes. A slight excess of carbon (∼5 %) was intentionally added to achieve critical temperatures above 10 K somewhat higher than reported before [12, 17]. Since the starting materials and the products are very sensitive to moisture all handling of materials was carried out in a glove box filled with Ar atmosphere. Samples were characterized by x-ray powder diffraction using a STOE diffractometer.

To determine the pressure dependence of $T_c$ two different techniques were employed: For pressures up to 1 GPa a sample of ∼1 mg $Y_2C_2I_2$ was mounted in a piston-type Cu-Be pressure cell [18] using Fluorinert as pressure medium. The static susceptibility of the cell and the sample was measured with an MPMS magnetometer in a magnetic field of 0.5 mT. For pressures up to 2 GPa a Cu-Be piston-type cell was employed and ac-susceptibility (field amplitude <1 Oe) on samples of approximately 2 mg was determined with a mutual inductance setup using the Quantum Design PPMS ac-resistivity option.

The pressures were determined from simultaneously determining $T_c$ of a small splinter of high purity Pb placed next to the sample [19].

Neutron powder diffraction at room temperature was carried out using the POLARIS time-of-flight diffractometer at the ISIS Facility, Rutherford Appleton Laboratory, using a Paris-Edinburgh type pressure cell (sample volume $\sim 100$ mm$^3$, $P \leq 5$ GPa). Carefully dried Fluorinert was used as pressure transmitting medium. Pressures were determined from the lattice parameters of NaCl added as an internal standard [20]. The pressure cell was placed in an evacuated sample tank in order to avoid background scattering from air and deterioration of the sample during the experiment. Multiphase profile refinements were carried out using the GSAS program package [21].

## 3. Results and Discussion

### 3.1. Pressure Dependence of $T_c$

As has been observed before, $T_c$ of $Y_2C_2I_2$ shows a monotonic *increase* with pressure (Fig.2). At low pressure, $T_c$ grows linearly at a rate of 1.4(1) K/GPa somewhat larger than reported previously [12, 22]. Towards high pressure the increase levels off and $T_c$ reaches a plateau value of $\sim 11.8$K. In the investigated pressure regime up to 2 GPa the pressure dependence of $T_c$ can well be described by a 2nd order polynomial, $T_c(P) = 10.15(3)$ K $+ 1.4(1)$ K GPa$^{-1} \times P$ - $0.31(4)$ K GPa$^{-2} \times P^2$. No hysteresis of $T_c$ was observed for pressures up to 2 GPa.

### 3.2. Pressure Dependence of the Crystal Structure

To establish the room-temperature equation of state (EOS) the cell parameters were determined from refinements of neutron powder patterns collected from a sample of $Y_2C_2I_2$ mixed with about 10 weight-% NaCl. In all, patterns were collected at seven different sample pressures obtained by progressively increasing the cell load up to a maximum of 70 tons. Unit cell volumes determined from x-ray powder diffraction measurements under pressure agree well with those from neutron powder diffraction results. From a fit of the cell volumes to a Birch-Murnaghan EOS [23, 24] we derived a bulk modulus of $B_0 = 37(1)$ GPa as described in detail in ref. [25].

The compressibilities of $Y_2C_2I_2$ are distinctly anisotropic as may be expected from the layered character of the structure. The largest relative decrease is observed along the $c$ direction with a compression of about 4% at 3 GPa whereas the relative compressibilities along $a$ and $b$ are more than a factor of two smaller. At small pressure values ($P \lesssim 0.5$ GPa) there is a small but perceivable nonlinearity in the decrease of the lattice parameters $b$ and $c$. Table 1 lists the refined lattice parameters, Fig. 4 displays the relative decrease of the cell parameters with pressure.

To follow the subtle changes of the structure under pressure and to gain structural parameters as input for the subsequent electronic structure calculations, we collected powder patterns obtained from a sample of $Y_2C_2I_2$ with extended exposure time ($\sim$15 h). The pressures were determined from a comparison of the cell volume with the EOS [25]. Fig. 3 displays three selected patterns taken at pressures up to $\sim$1.8 GPa in comparison with the calculated patterns obtained from the multiphase Rietveld profile refinements.

| $P$ (GPa) | $a$ (Å) | $b$ (Å) | $c$ (Å) | $\beta$ (°) | $V$ (Å$^3$) | techn. |
|---|---|---|---|---|---|---|
| 0.00 | 7.217(1) | 3.8814(7) | 10.435(2) | 93.37(1) | 291.80(9) | x-ray |
| 0.14 | 7.212(1) | 3.8765(6) | 10.411(2) | 93.29(2) | 290.57(6) | neutr |
| 0.72 | 7.1836(9) | 3.8635(1) | 10.320(1) | 93.34(1) | 285.93(4) | neutr |
| 1.27 | 7.153(1) | 3.8511(2) | 10.251(2) | 93.38(2) | 281.89(6) | neutr |
| 1.77 | 7.129(2) | 3.8398(2) | 10.190(2) | 93.44(2) | 278.46(8) | neutr |
| 3.03 | 7.064(4) | 3.818(3) | 10.019(2) | 93.58(4) | 270.43(16) | neutr |
| released | 7.213(1) | 3.8831(4) | 10.426(5) | 93.431(9) | 291.51(3) | neutr |

**Table 1.** Pressure dependence of the lattice parameters and the cell volumes of $Y_2C_2I_2$.

All these pure loading patterns showed Bragg reflections corresponding to $Y_2C_2I_2$ (SG $C2/m$) along with much weaker features associated with the pressure cell anvil material (WC, SG $P\bar{6}m2$) and Nickel, SG $Fm\bar{3}m$). The diffraction pattern of $Y_2C_2I_2$ was consistent with the space group and the cell parameters determined from previous x-ray and neutron powder diffraction data [22, 26, 27].

From the pattern emerging from $Y_2C_2I_2$ a total of 30 parameters (9 structural parameters) were refined. With increasing pressure the reflections broaden due to non-hydrostatic strain and the overall intensity decreases as a consequence of closing the gap between the anvils of the pressure cell. This is reflected in the relative fractions of the three phases obtained from the refinement. The refined fractions amounted to: 0.14 GPa: 84.7% / 13.9% / 1.4%; 0.72 GPa: 81.9% / 17% / 1.6%; and 1.77 GPa: 69.9% / 26.8% / 3.3% for $Y_2C_2I_2$, WC and Ni, respectively. The refined atom positional parameters are listed in Table 2.

In diffraction patterns collected at 3 GPa and 6 GPa there is evidence that a structural phase transition has taken place. Evidence for a structural phase transition in this pressure regime has also been seen in x-ray diffraction under pressure [28]. Release of pressure restores the low pressure phase completely. However, some reflections remain rather broad indicating significant residual strain left in the sample. The magnetic susceptibility measured on the pressure released sample reveal $T_c$ to be reduced to $\sim$ 7 K. $T_c$ is restored to its initial value of $\sim$ 10 K when the sample is reannealed at 1300 K for one day. x-ray diffraction patterns of this reannealed sample within error bars are identical to those of the initial sample.

Fig. 5 shows a nonlinear decrease of the C–C distance with pressure as calculated from the data given in Table 1 and 2 and a tilting of the C–C axis away from the apical Y metal atoms towards the triangular faces of the $Y_6$-octahedral cage. While we observe a smooth change of the angle with pressure, the C–C distance passes through a minimum

Structure and Electronic Properties of $Y_2C_2I_2$ under Pressure                                                        6| $P$ (GPa) | atom | x | y | z | $U_{\text{iso}}$ (Å$^2$) | $R_{wp}$ (%) |
|---|---|---|---|---|---|---|
| 0.14 | Y | 0.1429(12) | 0 | 0.1349(7) | 0.7(2) | 10.9 |
|  | C | 0.4227(13) | 0 | 0.0279(9) | 1.1(2) |  |
|  | I | 0.1726(15) | 0 | 0.6692(9) | 0.0(2) |  |
| 0.72 | Y | 0.1444(7) | 0 | 0.1350(4) | 1.8(1) | 6.2 |
|  | C | 0.4264(7) | 0 | 0.0298(5) | 2.1(2) |  |
|  | I | 0.1705(8) | 0 | 0.6657(6) | 0.9(1) |  |
| 1.77 | Y | 0.1461(13) | 0 | 0.1369(7) | 1.5(2) | 9.9 |
|  | C | 0.4277(12) | 0 | 0.0314(9) | 1.6(2) |  |
|  | I | 0.1714(17) | 0 | 0.6579(12) | 0.6(2) |  |

**Table 2.** Atomic positions of $Y_2C_2I_2$ at room temperature as refined from the powder patterns collected at the indicated pressures. All atoms are at the Wyckoff position $4i$ ( space group no. 12; $C2/m$). The rightmost column lists the weighted-profile $R$-factor.

and grows again towards 3 GPa. This behaviour heralds the pressure induced phase transition pressures above $\sim 3.5$ GPa. From the tendency of the C–C group to tilt away from the apical metal atoms towards the triangular faces of the metal atom octahedra which in turn enables increased bonding to three metal atoms one may expect that the phase transition will involve significant electronic and structural changes. The limited experimental accuracy of the refined atom parameters at 3.03 GPa, however, does not permit further confident conclusions at present.

## 4. Electronic Structure Calculations

All electronic structure calculations have been performed within the full-potential linearized plane wave (LAPW) method as implemented in the WIEN97 code [29]. Computational details were similar to those used in ref. [15]. For Brillouin-zone (BZ) integrations three different **k** meshes up to 5408 **k** points within the BZ were used and the results were checked for convergence. The lattice constants and, in a first step, also the atomic positions were taken from the neutron-diffraction results. In a subsequent iterative procedure the atomic positions were changed according to the calculated atomic forces until the force-free geometry and, hence, the total-energy minimum was reached.

The calculations confirm the double peak structure in the density of states (DOS) at the Fermi energy [15]. The peaks centered at about 70 meV above and at about 100 meV below the Fermi level emerge from C $p$ and Y $d$ bands of low dispersion along the $\Gamma$-$A$ ($A = (0,0,\bar{\frac{1}{2}})$) direction intersecting the Fermi level. This feature turns out to be particularly sensitive to subtle structural changes induced by applying pressure. As shown in Fig. 6 the pseudogap closes by applying pressure with a linear rate at low pressures (cf. inset) and reaches saturation above $\sim 2$ GPa. At small pressure values, the logarithmic derivative $dlnN(E_{\text{F}}, P)/dP$ amounts to $\sim 0.2$ GPa$^{-1}$. At 1.77 GPa the pseudogap completely vanishes and a broad peak at the Fermi level results. This behaviour is in contrast to a simplified rigid band picture for which one would expect the Fermi level to be upshifted across the double peak feature in the DOS.

For the quantitative discussion of the pressure dependence of $T_c$ we start from the Allen-Dynes modification of the McMillan equation [30, 31]. Neglecting any pressure dependence of the Morel-Anderson effective Coulomb pseudopotential $\mu^*$ [32] and using approximations suggested by Loa *et al.* [6] one arrives at

$$\frac{d\,lnT_c}{dP} \approx C\,\frac{d\,lnN(E_F)}{dP} + (1 - 2C)\frac{\gamma}{B_0} \qquad (1)$$

with $C = (1.04 + 0.4\mu^*)\lambda/(\lambda - \mu^*(1 + 0.62\lambda))^2$ and $\lambda$ being the electron-phonon coupling constant $\lambda = N(E_F) <I^2>/m<\omega^2>$. $<I^2>$ is the square of the matrix element of the electron-phonon interaction averaged over the Fermi surface, $m$ the atomic mass, and $<\omega^2>$ is an average over the phonon spectrum defined explicitly in ref. [30].

With $\lambda \sim 1$ as derived from heat capacity measurements [33], assuming $\mu^* \sim 0.1$ ($C \sim 1$), a mode Grüneisen parameter $\gamma \sim 1$–$2$ and using the bulk modulus $B_0 = 37(1)$ GPa, the experimental observation $1/T_c(0) \cdot dT_c/dP = 0.14(1)$ GPa$^{-1}$ suggests an increase of the electronic density of states with pressure, $1/N(E_F) \cdot dN(E_F)/dP$, of 0.17–0.20 GPa$^{-1}$. This value is in very good agreement with the initial slope of the pressure increase of $N(E_F)$ (see inset Fig. 6). A detailed analysis revealed an increase of all partial DOSs with pressure. We could not identify a particular partial DOS which dominates the increase, but all states similarly contribute to this pressure-induced enhancement.

For Y$_2$C$_2$Br$_2$, Y$_2$C$_2$Br$_1$I$_1$ and Y$_2$C$_2$Br$_{0.5}$I$_{1.5}$, we observed a relative decrease of $T_c$ with pressure between -2.6% and -9.8% GPa$^{-1}$ which is about the order of magnitude one would expect due to lattice stiffening alone [12]. Y$_2$C$_2$I$_2$ and La$_2$C$_2$Br$_2$ (+18% GPa$^{-1}$, [12]) appear to have a special electronic situation which favours an increase in $T_c$ with the application of external pressure.

## 5. Summary

We have analysed the pressure-induced changes of the crystal structure of the layered superconductor Y$_2$C$_2$I$_2$ by a detailed neutron diffraction study under pressure. The refined atomic parameters enable us to relate the observed increase of $T_c$ with pressure to pressure-induced changes in the electronic structure. Good quantitative agreement between experiment and theoretical predictions is observed.

**Acknowledgement** Useful discussions with O. Dolgov, J. Köhler, J. Kortus, I. Loa and K. Syassen are gratefully acknowledged. We thank E. Brücher, H. Eckstein, S. Höhn, G. Siegle and D. J. Francis for technical assistance. CAD appreciates support from the Austrian Science Fund, project P13430.

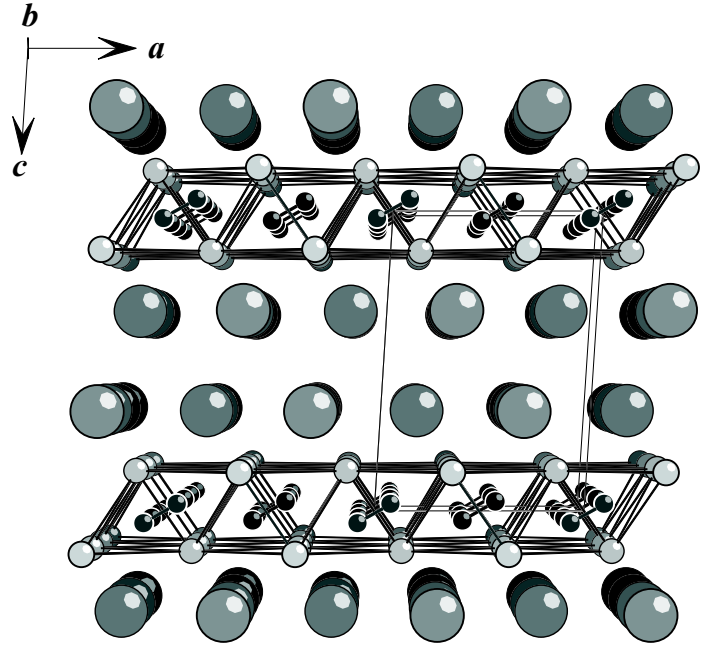

**Figure 1.** Crystal structure of $Y_2C_2I_2$ with a unit cell outlined. I, Y, and C are drawn with decreasing size.

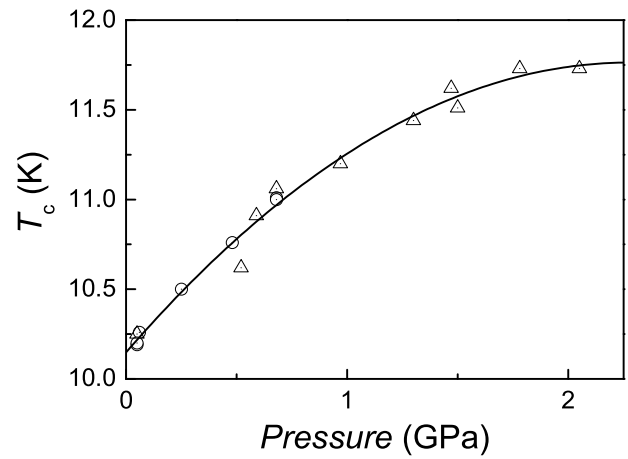

**Figure 2.** Variation of $T_c$ of $Y_2C_2I_2$ with pressure. The full line corresponds to a polynomial of 2nd degree with parameters given in the text. $T_c$'s have been determined from static susceptibility (○) and ac susceptibility (△) measurements.

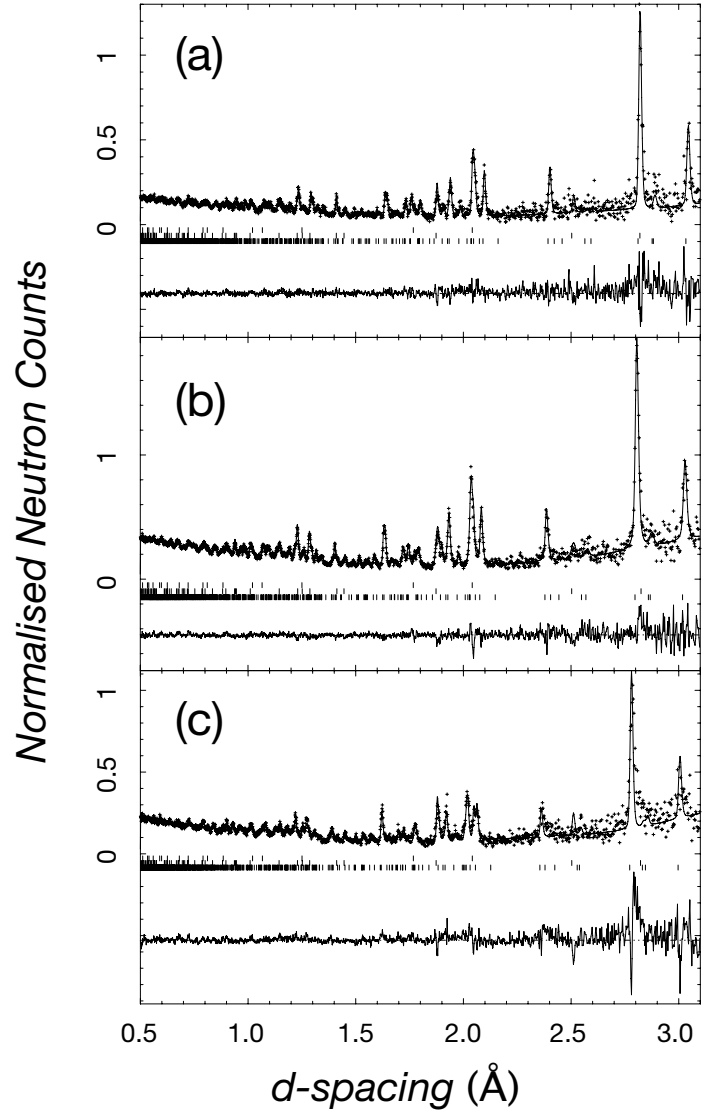

**Figure 3.** Neutron powder diffraction patterns of $Y_2C_2I_2$ at (a) 0.14 GPA, (b) 0.72 GPa and (c) 1.77 GPa. Data collected at $2\theta = 90^o$ are shown as dots. The solid lines are the results of Rietveld profile refinements. Vertical bars indicate the position of the reflections used to generate the patterns. Upper bar row, medium bar row and lower bar row indicate the reflection of Ni, WC (from the anvils) and of the $Y_2C_2I_2$ sample, respectively. The lower solid lines represent the difference between the observed and calculated patterns.

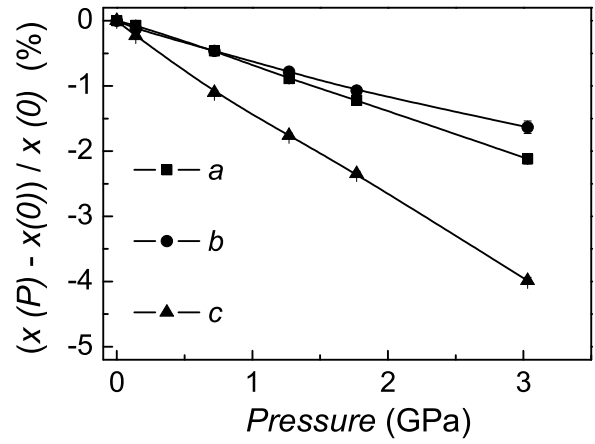

**Figure 4.** Relative change of the lattice parameters *a*, *b*, *c* with pressure. Size of the symbols indicates error bars. Full lines are guides to the eye.

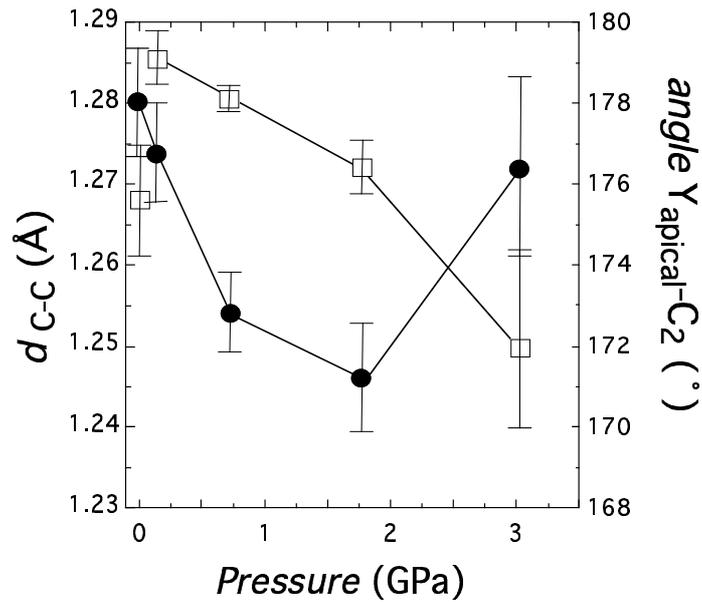

**Figure 5.** C–C distances (●) and tilting angle (□) of the axis of the C–C dumbbell with respect to the apical Y metal atoms.

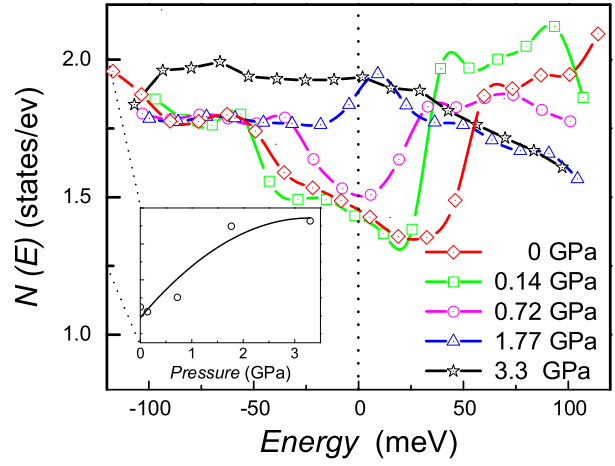

**Figure 6.** Electronic density of states $N(E)$ in the close neighbourhood of the Fermi level obtained from the total energy optimized crystal structure. The inset shows the pressure dependence of $N(E_F)$. The full line is a guide to the eye.